# Position dependent effective mass and pseudo-hermetic generator in Hamiltonians with *PT* symmetry

F. Soliemani; Z. Bakhshi *

*Physics Department, Shahed University, Tehran, Iran*

**Abstract** In this paper we present a general method to solve non hermetic potentials with PT symmetry using the introduction of two first-order operator against $\eta$-pseudo-hermetic($\eta$-weak-pseudo-hermiticity) with position dependent effective mass and applying it to Dirac equation with spinor wave function and non-hermitic potentials with real energy spectrum considering position dependent effective mass. Also Pöschl-teller complex and Eckart potentials will be discussed as examples.

## I. INTRODUCTION

For the past few years, studying quantum systems considering position dependent effective mass (*PDEM*) has been taken into consideration. Quantum particles with position dependent effective mass create interesting and useful methods to study physics. These methods ae typically used to study energy density in multiple subjects issues, Determination of the electronic properties of multiple semiconductor structures and also describing the properties of multiple bonds and quantum points [1-3].

Non-hermetic models have several usages in studying physics systems such as nuclear Physics, quantum field theories, etc. we have used different methods to discuss the solution of relativistic and non-relativistic equation with non-hermitic Hamiltonian with real or imaginary energy spectrum. Therefore solving Dirac equation with effective mass for an imaginary potential could be an interesting discussion [4-6].

Discussing non-hermit Hamiltonian with *PT* symmetry reduces Hamiltonian's hermitage requirement as an essential situation for real energy spectrum. Non-hermit Hamiltonian is true when $\hat{O}H\hat{O}^{-1} = \hat{O}^{-1}H\hat{O} = H$ in which $\hat{O} = PT$ is a combined operator of Parity and time inversion transformations. Thus they are attributed to PT symmetry. If a potential is $v(x) = v^*(x)$ sub-conversion $i \rightarrow -i$ and $x \rightarrow -x$, we consider it a potential with *PT* symmetry [7-8].

Currently a more generalized method is presented to verify if an energy spectrum is true and it suggests a *PT* symmetrical Hamiltonian consists of sub-systems named pseudo-hermitic. Which means *H* Hamiltonian is pseudo-Hamiltonian if it follows similarity conversions below:

$\hat{\eta}H\hat{\eta}^{\dagger} = \hat{\eta}^{\dagger}H\hat{\eta} = H$

in which (†) is an accessory marker and $\eta$ is a hermetic inverse linear operator and $\eta = O^{\dagger}O$, $O: \mathcal{H} \rightarrow \mathcal{H}$ ($\mathcal{H}$ is Hilbert Space) [9-10].

Therefore we can claim that neither hermitic nor *PT* symmetry is the necessary requisite to verify if a quantum Hamiltonian is true.

Considering some weak theories and by not bounding $\eta$ to the hermitic, linear or reversible requisite, we can write *H* as a week- pseudo-Hermitian.

In this paper, we study one dimensional non-hermitic imaginary and non-imaginary with true energy spectrum in Dirac`s equation in context of position dependent effective mass. Here, Dirac equation will be solvable by reformulating to a Schrödinger-like equation with position dependent effective mass by introducing first-order cross-operator–week-pseudo-hermitian for creating a class of *PDEM* Hamiltonians in the obtained mass function.

In explaining this article, we will act as described below.

---

* Corresponding author: z.bakhshi@shahed.ac.ir

In section 2, Dirac equation within position dependent effective mass is discussed in an external magnetic field.

In section 3, two first-order hermitic and non-hermitic operators with position dependent effective mass will be introduced.

In section 4, we come up with a general formula for $\eta$-week-pseudo-Hermitian using the change of focal points on Hamiltonian formerly resulted from relativistic Dirac equation and comparing it with non-relativistic Schrödinger equation.

In section 5 and 6, to clarify the problem, we use Pöschl-teller complex and Eckart potentials.
In the last section we will have conclusions.

## II. One dimensional Dirac equation with position dependent effective mass

Relativistic Dirac equation with position dependent effective $M(x) = m_0 m(x)$ for a moving particle in an external electromagnetic field $A_\mu$ is:

$$[i\gamma\mu(\partial\mu + ieA\mu)M(x)]\psi = 0 \; ; \qquad \hbar = C = 1 \tag{1}$$

in which $\gamma_\mu (\mu = 0,1,2,3)$, the gamma matrix is written as follows:

$$\gamma 0 = \begin{pmatrix} 0 & I \\ I & 0 \end{pmatrix} \; ; \quad \gamma i = \begin{pmatrix} 0 & -\sigma^i \\ \sigma^i & 0 \end{pmatrix} \tag{2}$$

I matrix are the unit $2 \times 2$ matrix and $\sigma^i$ is the Pauli matrix. When Vector potential doesn`t attend and with considering $\upsilon(x) = eA_0(x)$ one dimensional Dirac equation in $\psi(x,t) = e^{-i\varepsilon t}$, $\psi(x)$ in the pairing plan for up component is $\varphi(x)$ and for down component is $\theta(x)$ in spinor wave function $\psi(x)$ is written as:

$$\left[i\frac{d}{dx}\begin{pmatrix} 0 & -1 \\ 1 & 0 \end{pmatrix} + (\varepsilon - \upsilon(x))\begin{pmatrix} 0 & 1 \\ 1 & 0 \end{pmatrix} - M\begin{pmatrix} 1 & 0 \\ 0 & 1 \end{pmatrix}\right]\begin{pmatrix} \varphi(x) \\ \theta(x) \end{pmatrix} = 0 \tag{3}$$

The written equation can be re-written as:

$$-i\frac{d\theta(x)}{dx} + (\varepsilon - \upsilon(x))\theta(x) - M(x)\varphi(x) = 0 \tag{4}$$

$$i\frac{d\varphi(x)}{dx} + (\varepsilon - \upsilon(x))\varphi(x) - M(x)\theta(x) = 0 \tag{5}$$

By eliminating the bottom spinor factor in $\theta(x)$ two above equations and combining these equations, we can write a Schrödinger-like equation for up spinor factor $\varphi(x)$ [11]:

$$-i\frac{d^2\varphi(x)}{dx^2} + \left[2\varepsilon\upsilon(x) - \upsilon^2(x) - i\frac{d\upsilon(x)}{dx} - i\frac{1}{M(x)}\frac{dM(x)}{dx}(\varepsilon - \upsilon(x))\right]\varphi(x) + \frac{1}{M(x)}\frac{dM(x)}{dx}\frac{d\varphi(x)}{dx} = (\varepsilon^2 - M^2(x))\varphi(x) \tag{6}$$

### III. η-pseudo-hermitic

Two first-order hermitic and non-hermitic operators with effective mass will be introduced as below [12-15].

$$\eta_1 = -i\left[\mu(x)\frac{d}{dx}\right] + F(x) \tag{7}$$

$$\eta_2 = \mu(x)\frac{d}{dx} + iF(x) \tag{8}$$

In which, $\mu(x) = \frac{1}{M(x)}$ and $M(x) = m_0 m(x)$ $x \in (-\infty, +\infty)$ is position dependent effective mass.

$$\eta_j H = H^+ \eta_j \quad ; \quad j = 1,2 \tag{9}$$

Naming $W(x) = \mu(x)\frac{d}{dx}[i(\mu(x)(\varepsilon - \upsilon(x))]$ and applying equation (9) in equation (6) written equations will be resulted:

$$2iW(x)\mu(x) = -i2\mu(x)^2 F'(x) \tag{10}$$

$$W(x) = -\mu(x)F'(x) \tag{11}$$

$$-i\mu^2(x)F''(x) - i\mu(x)\mu'(x)F'(x) = i\mu(x)W'(x) \tag{12}$$

$$-V'(x) - W(x)F(x) = W(x)F(x) \tag{13}$$

which leads to:

$$V_j(x) = -F^2(x) - i\mu(x)F'(x) + \alpha_0 \tag{14}$$

And $\alpha_0 \epsilon R$ is integration constant, therefore:

$$H = -\mu(x)^2 \partial_x^2 - \mu(x)\mu(x)'\partial_x + V_j(x) \tag{15}$$

### IV. Canonical transformations

In this section by canonical transformations as below:

$$\varphi(x) = \psi(q(x)) \tag{16}$$

We will have:

$$-\mu(x)^2 \left[\frac{d^2\psi(q)}{dq^2}(q'(x))^2 + \frac{d\psi(q)}{dq}q''(x)\right] - \mu(x)\mu(x)'\left[\frac{d\psi(q)}{dq}q'(x)\right] + (-F^2(x) - \mu(x)F'(x) + \alpha_0 - E)\psi(q(x)) = 0 \tag{17}$$

Comparing above equation with non-relativistic Schrödinger equation and placing $q' = \frac{1}{\mu(x)}$ in equation (17), we will have a $V_{eff}(x)$ general formula like equation (20):

$$-\frac{d^2\psi(q)}{dq^2} + (-F^2(q(x)) - \mu(x)F'(x) + \alpha_0 - E)\psi(q(x)) = 0 \tag{18}$$

$$\frac{dF_j(x)}{dx} = \frac{dF_j(q)}{dx} = \frac{dq}{dx}\frac{dF_j(q)}{dq} = \frac{1}{\mu}\frac{dF_j(q)}{dq} \tag{19}$$

$$V_{eff}(q) = -F_j^2(q) - F_j'(q) + \alpha_0 \tag{20}$$

**Examples:**

We consider two cases as follows:

**Example1: Pöschl-teller complex potential**

As an example, we consider a $\eta$-pseudo-hermitic with *PT* symmetry, Pöschl-teller is studied in [16] using complex Lie algebra:

$$V(x) = V_1 \operatorname{cosech}^2(\alpha x) - V_2 \operatorname{cosech}(\alpha x)\coth(\alpha x) \tag{21}$$

where $T = x - c - i\gamma$, $V_1 > -\frac{1}{4}$ and $V_2 \neq 0$.

$$E_{n,\varepsilon} = -\left[\frac{1}{2}\right]\left[\sqrt{V_1 + \frac{1}{4} + |V_2|} + \varepsilon\sqrt{V_1 + \frac{1}{4} - |V_2|} - n - \frac{1}{2}\right]^2 \quad \varepsilon = \pm 1 \tag{22}$$

$$V_{eff} = -V_2^2 \operatorname{cosech}^2(\alpha x) - iV_2 \operatorname{cosech}(\alpha x)\coth(\alpha x) \tag{23}$$

In other words, $\eta$-pseudo-hermitic operator will be written as:

$$F_{(q)} = V_2 \operatorname{cosech}(q) \rightarrow F'_{(q)} = V_2 \operatorname{cosech}(q)\coth(q) \tag{24}$$

$$E_{n,\varepsilon+1} = \left[|V_2| - n - \frac{1}{2}\right]^2 \quad n = 0,1,2,\ldots \quad n_{max} < \left(|V_2| - \frac{1}{2}\right) \tag{25}$$

And for $|V_2| < \frac{1}{2}$, $\varepsilon = \pm 1$ and for $|V_2| > \frac{1}{2}$, $\varepsilon = -1$ we have empty stages of energy levels and in such situation the phenomenon of crossing the energy level happens [17-19]. Considering $f(x) \in R$ and $q(x) = \pm \ln f(x)$ a class of $\eta$-pseudo-hermitic Pöschl-teller is resulted:

$$M(x) = [\pm \partial_x \ln f(x)] \rightarrow f(x) = exp(\pm \int M(z))dz \tag{26}$$

$$V_{eff} = -4V_2^2 \frac{f(X)^2}{(f(X)^2-1)^2} - 2iV_2 \frac{f(X)((f(X)^2+1)}{(f(X)^2-1)^2} \tag{27}$$

**Example2: Eckart shape invariant potential:**

As another example, we discuss Eckart shape invariant shape potential [20].

$$V(x) = A^2 + \frac{B^2}{A^2} + A(A-1)\,cosech^2(q) - 2B\,coth(q) \tag{28}$$

$$E = (A+n)^2 - A^2 + \frac{B^2}{A^2} - B^2(A-n)^2 \tag{29}$$

With considering $F_j(q)$ as follows:

$$F_j(q) = A\,coth(q) + \frac{B}{A} \qquad B > A^2 \tag{30}$$

We will have:

$$V_{eff}(x) = A^2 + \frac{B^2}{A^2} + A(A-i)\,cosech^2(q) - 2B\,coth(q) \tag{31}$$

Considering the fact that there are many options for $\eta$-pseudo-hermitic Eckart potential, $q(x)$ is written as $ln(x^2 - 1)$:

$$q(x) = \pm ln(x^2 - 1) = \int M(z)dz \tag{32}$$

Thus, we will have:

$$V_{eff}(x) = A^2 + \frac{B^2}{A^2} + A(A-i)\,\frac{(x^2+1)^2}{((x^2+1)^2-1)^2} - 2B\,\frac{(x^2+1)^2+1}{(x^2+1)^2-1} \tag{33}$$

**V. Conclusion**

Based on previous studies, we considered a more generalized situation for non-hermitic Hamiltonian so that *PT* symmetry Hamiltonians consists of sub-system named $\eta$-pseudo-hermitic, if they follow $\eta H = H^+\eta$ similarity conversion. In this paper we introduced two first-order hermitic and non-hermitic operators and discussed applying it on Schrödinger-like resulted from Dirac equation with effective mass, then we came up with a general formula in forms of $V_j(x) = F_j^2(q) - F'_j(q) + \alpha_0$ and we could it to solve potentials with *PT* symmetry. The study of the imaginary potential of Pöschl-teller revealed the effect of energy loss.

**References**

[1] C. Quesne, Ann. Phys. 321 (2006) 1221.
[2] T. Tanaka, J. Phys. A; Math and Gen 39 (2006) 219.
[3] O. Mustafa and S. H. Mazharimousavi, Phys. Lett. A 358 (2006) 259.
[4] C. M. Bender, H. F. Jones and R. J. Rivers, Phys. Lett. B 625 (2005) 333.
[5] C. S. Jia and A. de Souza Dutra, J. Phys. A: Math. Gen. 39 (2006) 11877.
[6] Z. Ahmed, Phys. Lett. A 286 (2001) 2316.
[7] C. M. Bender, S. Boettcher and P. N. Meisinger, J. Math. Phys. 40 (1999) 2201.
[8] O. Mustafa, J. Phys. A: Math. Gen. 36 (2003) 5067.
[9] A. Sinha and P. Roy, Czech. J. Phys. 54 (2004) 129.
[10] O. Mustafa and S. H. Mazharimousavi, Czech. J. Phys. 56 (2006) 967.
[11] H. Panahi and Z. Bakhshi, Acta. Physica. Polonica B 41 (2010).
[12] O. Muatafa, S Habib Mazharimousavi, (2008).
[13] A. Mostafazadeh, J. Math. Phys. 43 (2002) 2814.
[14] A. Mostafazadeh, Nucl. Phys. B 640 (2002) 419.
[15] A. Mostafazadeh, J. Phys. A: Math. Gen. 38 (2005) 3213.
[16] B. Bagchi and C. Quesne, Phys. Lett. A 300 (2002) 173.
[17] R. Kretschmer and L. Szymanowski, Czech. J. Phys 54 (2004) 71.
[18] O. Mustafa and M. Znojil, J. Phys. A: Math. Gen. 35 (2002) 8929.
[19] B. F. Samsonov and P. Roy, J. Phys. A: Math. Gen. 38 (2005) L24.
[20] G. Levai, J. Phys. 22 (1989)703.